# 40% boost in extreme ultraviolet conversion efficiency via simultaneous dual-beam 2-μm laser irradiation

NAOKI NAGAHAMA,[1†] KAITO NISHIMIYA,[2†] SHUNYA YAMAMOTO,[1] HAYATO YAZAWA,[1] YUTA TAKAI,[1] CHISATO TANAKA,[1] KAZUYUKI SAKAUE,[3] ATSUSHI SUNAHARA,[4] GERRY O'SULLIVAN,[5] SHINICHI NAMBA,[6] TAKESHI HIGASHIGUCHI,[1,*] AND EIJI J. TAKAHASHI[2,**]

[1]*Department of Electrical and Electronic Engineering, Faculty of Engineering, Utsunomiya University, 7-1-2 Yoto, Utsunomiya, Tochigi 321-8585 Japan*
[2]*Extreme Photonics Research Group, RIKEN Center for Advanced Photonics, RIKEN, 2-1 Hirosawa, Wako, Saitama 351-0198, Japan*
[3]*Nuclear Professional School, School of Engineering, The University of Tokyo, 7-3-1 Hongo, Bunkyo, Tokyo 113-8656, Japan*
[4]*Institute of Multidisciplinary Research for Advanced Materials, Tohoku University, 2-1-1, Katahira, Aoba-ku, Sendai 980-8577, Japan*
[5]*School of Physics, University College Dublin, Belfield, Dublin 4, Ireland*
[6]*Department of Advanced Science and Engineering, Hiroshima University, 1-4-1 Kagamiyama, Higashihiroshima, Hiroshima 739-8527, Japan*
[†]*These authors contributed equally to this work.*
**higashi@a.utsunomiya-u.ac.jp*
***ejtak@riken.jp*

**Scaling extreme ultraviolet (EUV) source power for next-generation lithography demands higher conversion efficiency (CE) at reduced per-pulse energies. We demonstrated a 40% CE enhancement by simultaneous dual-beam irradiation of a planar Sn target with a 2090-nm, 20-ns Ho:YAG laser. Single-beam irradiation at 40 mJ yielded an EUV CE of 2.6%; splitting the same total energy equally into two beams of 20 mJ each — at identical peak intensity — raised the EUV CE to 3.6%, which was the highest reported for 2-μm-driven laser-produced plasma sources. The EUV source size (60–70 μm) and energetic-ion spectra were nearly identical across both configurations, confirming comparable plasma conditions. Because the scheme requires only passive beam splitting and scales readily to three or more beams, it offers a practical route toward multi-kW-class, energy-efficient EUV sources for high-NA and hyper-NA lithography.**

Solid-state lasers operating at a wavelength of 2 μm are attractive drivers for laser-produced plasma (LPP) extreme ultraviolet (EUV) sources: they combine high in-band (2% bandwidth at 13.5 nm) conversion efficiency (CE) [1-3] with a wall-plug efficiency of 10%–15%, far exceeding the ~ 3% of $CO_2$ lasers whose total power consumption approaches 1 MW [4]. However, the average power and per-pulse energy demonstrated to date with 2-μm lasers — 270 W at 10 kHz for a Ho:YAG master oscillator power amplifier (MOPA) [5] — remain far below the multi-kW level needed for high-volume EUV lithography at 100 kHz [6]. Scaling is fundamentally constrained by thermal damage to the gain medium and surrounding optics. Multiple-beam irradiation is a proven strategy to enhance the EUV CE when the per-beam pulse energy is limited; at 1 μm, it has raised the EUV CE to 4.7% [7]. As yet, no experimental demonstration of multiple-beam irradiation using a 2-μm laser has been reported. The extension to 2 μm is non-trivial: the fourfold decrease in critical density shifts the laser absorption zone further from the target surface, fundamentally altering the plasma heating profile and making the outcome of multi-beam irradiation difficult to predict from 1-μm results alone.

The industrial demand for higher EUV source power continues to grow. Semiconductor technology nodes are advancing into the angstrom regime, and the numerical aperture (NA) of EUV exposure optics is transitioning from 0.33 to 0.55 (high-NA), with scaling to hyper-NA (0.75) under active research and development [8]. An in-band EUV power of 1 kW at intermediate focus has recently been demonstrated at 100 kHz using a $CO_2$-laser-driven Sn-droplet source with an EUV CE of 6% – 7% [6], but the EUV exposure tools (EXE:5000 series) consume up to 33,600 kWh/day, making both higher EUV CE and improved drive laser efficiency essential [4]. These constraints motivate the exploration of 2-μm solid-state drivers, for which several proof-of-principle single-beam experiments have confirmed promising EUV CE values [1-3,9,10].

Achieving high EUV CE at per-pulse energies on the order of 10 mJ poses specific difficulties related to plasma confinement and energy deposition. Maintaining sufficient laser intensity at low energies requires tighter focusing, which reduces the EUV-emitting area. Furthermore, three-dimensional plasma expansion causes rapid cooling that shortens the emission duration and limits the total EUV yield [11]. Multiple-beam irradiation mitigates these effects by broadening the laser-heated region, thereby enlarging the EUV-emitting volume and surface area. This strategy is particularly advantageous when the output of high-repetition-rate laser systems is constrained by thermal damage — the very conditions facing 2-μm driver development. We note that the highest EUV CE reported for 2-μm-driven LPP sources to date is 4%, obtained from mass-limited Sn microdroplets [3]. In practical high-repetition-rate EUV sources, however, the droplet is first deformed into a disk-shaped "pancake" target by a nanosecond pre-pulse before the main pulse arrives [1-3,6]. A planar Sn target closely approximates this pancake geometry, making planar target experiments the most relevant benchmark for evaluating the EUV CE enhancement achievable under realistic source conditions. A planar Sn target directly approximates this pancake geometry in terms of the target aspect ratio and the laser–target interaction angle, although the initial density profile of a pre-pulse-expanded droplet differs from that of a bulk solid surface. Nevertheless, planar-target experiments remain the most accessible and reproducible benchmark for evaluating multi-beam EUV CE enhancement under controlled conditions.

In this Letter, we demonstrate a 40% enhancement in EUV CE by simultaneously irradiating a planar Sn target with two mid-infrared laser pulses (2090 nm, Gaussian profile, 20 ns FWHM). Single-beam irradiation at 40 mJ yielded an EUV CE of 2.6%; splitting the same total energy into two equal beams of 20 mJ each — at identical peak laser intensity — raised the EUV CE to 3.6%, the highest value reported for any planar-target experiment with a 2-μm driver and narrows the gap with the 4% EUV CE obtained from optimized mass-limited droplet targets [3], despite the use of a planar geometry with inherently strong re-absorption losses. We also examine, through EUV source-size measurements and high-energy ion spectra, whether comparable plasma-formation conditions are established in both configurations, thereby supporting the applicability of the dual-beam scheme to next-generation high-power EUV sources.

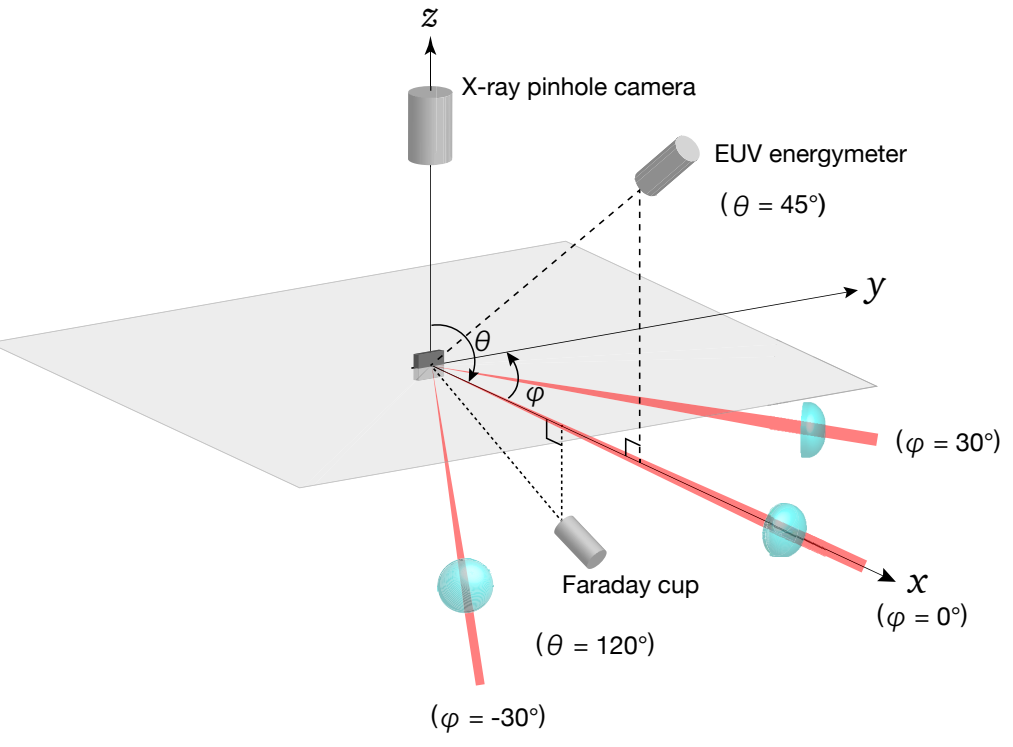

Fig. 1. Schematic diagram of the experimental apparatus.

A schematic diagram of the experimental apparatus is shown in Fig. 1. The experiment was performed using a $Q$-switched Ho:YAG laser system at a laser wavelength of 2090 nm pumped by a Tm-based laser. The laser pulse duration was 20 ns [full width at half-maximum (FWHM)] at a repetition rate of 1 kHz. The pulse energy fluctuations were approximately 1%. We measured the focal spot diameter by the Knife-edge method. The error of the focal spot diameter was evaluated to be 10%. The maximum pulse energy was 50 mJ, and we varied the number of laser beams, keeping the laser intensity in each at $(1-2)\times 10^{11}$ W/cm$^2$, which corresponds to the optimum laser intensity for the EUV emission. The EUV emission at 13.5 nm originates from highly charged ions ranging from $Sn^{7+}$ to $Sn^{14+}$ [12,13]. The optimal electron temperature is 30 eV [13,-15], as determined by a collisional radiative (CR) model [16,17]. At a laser wavelength of 2 μm, the optimum laser intensity is expected to be $(0.8-2)\times 10^{11}$ W/cm$^2$ to achieve an electron temperature of about 30 eV and the maximum EUV CE. Therefore, each laser beam should be tuned to the optimum laser intensity by changing the focusing lens positions and focal spot diameters.

The EUV energy meter, which consisted of a concave Mo/Si mirror, a thin-foil Zr filter, and a Si-based x-ray diode, recorded the emission energy which was used to evaluate the EUV CE. The EUV energy meter was used to evaluate the in-band (2% BW) EUV CE at 13.5 nm with the EUV energy meter positioned at 45° relative to the incident laser axis at 0°. The EUV CE values reported here assume isotropic emission over 2π sr, as is standard for planar-target experiments [1]. An EUV pinhole camera captured the source images ~~in Fig. 3~~ with a thermoelectrically cooled, back-illuminated x-ray charge-coupled device (CCD) (Andor Technology). We positioned the EUV pinhole camera at 90° with respect to the incident laser axis at 0°. In addition, the Faraday cup, positioned at 30°, was used to measure the time-of-flight (TOF) signal from the fast ions. From the TOF measurements, we obtained the charge-integrated ion energy spectra.

Before investigating the effect of dual-beam irradiation, the EUV CE and spectral characteristics were measured under single-beam conditions and benchmarked against the radiation-hydrodynamic simulation code Star-2D. Figure 2 shows the laser-intensity dependence of the EUV CE and spectral purity (SP). The EUV CE is defined as the ratio of the in-band EUV emission energy to the incident laser energy, where the in-band emission is measured at a center wavelength of 13.5 nm within a 2% bandwidth, over a solid angle of 2π sr. The EUV SP is defined as the fraction of in-band emission relative to the total emission measured over the wavelength region of 6.4 – 22 nm.

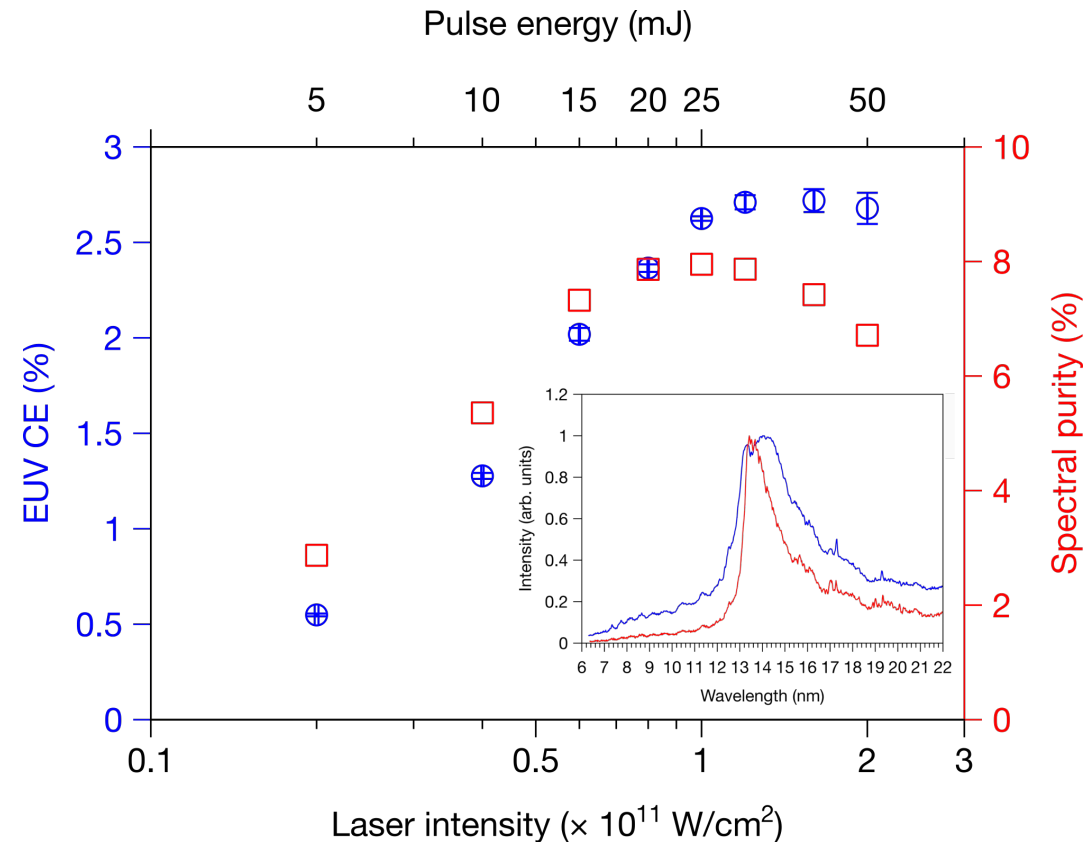

Fig. 2. Laser intensity dependence of the EUV CE for the single beam irradiation. The inset shows the observed EUV spectra for two different laser wavelengths at 1064 (Blue) and 2090 nm (red).

The EUV CE peaked at 2.6% at a laser intensity of $1.6\times 10^{11}$ W/cm$^2$, while the EUV SP reached a maximum of 8% at $1\times 10^{11}$ W/cm$^2$. Both values are consistent with previously reported results [1]. Because the EUV CE is the product of the laser absorption efficiency, the radiative conversion efficiency, and the SP, the slight offset between the EUV CE- and SP-optimal intensities lies within experimental uncertainty and does not affect the overall interpretation. According to the EUV spectra in the inset in Fig. 1, the SPs were observed to be 4% for 1064 nm and 8% for 2090 nm. Compared with the 1-µm laser-produced plasmas, the 2-µm driver yields higher EUV SP and stronger suppression of the out-of-band emission in the inset in Fig. 1. This improvement is attributed to the fourfold reduction in critical electron density at the laser wavelength of 2 µm, which produces an optically thin plasma, thereby narrowing the emission to the 13.5 nm band.

To quantify the underlying plasma parameters, the radiation-hydrodynamic simulation of "Star-2D" [18,19] was performed at the EUV CE-optimal laser intensity under present experimental conditions. The calculated EUV CE of about 3.2% essentially reproduces the experimental value, and the simulated spectrum closely matches the measured spectrum. The EUV-emitting region was calculated to have an electron temperature of 30 eV and an electron density of $1 \times 10^{20}$ cm$^{-3}$, consistent with optimizing the ionic charge-state distribution responsible for 13.5 nm emission.

The single-beam measurements establish a reliable baseline for the planar Sn target geometry — a direct analog of the so-called pancake target used in high-repetition-rate sources, as discussed above. Building on this baseline, we show the EUV CE, emission images, and energetic-ion spectra under simultaneous dual-beam irradiation to assess the potential of the multi-beam scheme to further enhance EUV CE.

The Ho:YAG laser output was split into two equal beams by a 50:50 beam splitter, with each beam's pulse energy set to 20 mJ through careful power balancing of the two arms. The combined pulse energy of 40 mJ matched that used in the single-beam experiments, thus keeping the total laser energy incident on the planar Sn target constant. Because each beam delivered half the total energy, the focusing lens in each arm was repositioned to maintain the optimal laser intensity of $1.6 \times 10^{11}$ W/cm$^2$, identical to the single-beam value. The resulting focal spot diameter was 30 µm, slightly smaller than the 40-µm diameter obtained in the single-beam case. The EUV signals recorded by the EUV energy meter are shown in Fig. 3. Despite the identical total incident energy, the dual-beam configuration produced a substantially higher EUV signal than the single-beam case. The EUV emission energy, obtained by time-integrating the calorimeter waveform, increased by 40%, raising the EUV CE from 2.6% to 3.6% — a 1.4-fold enhancement.

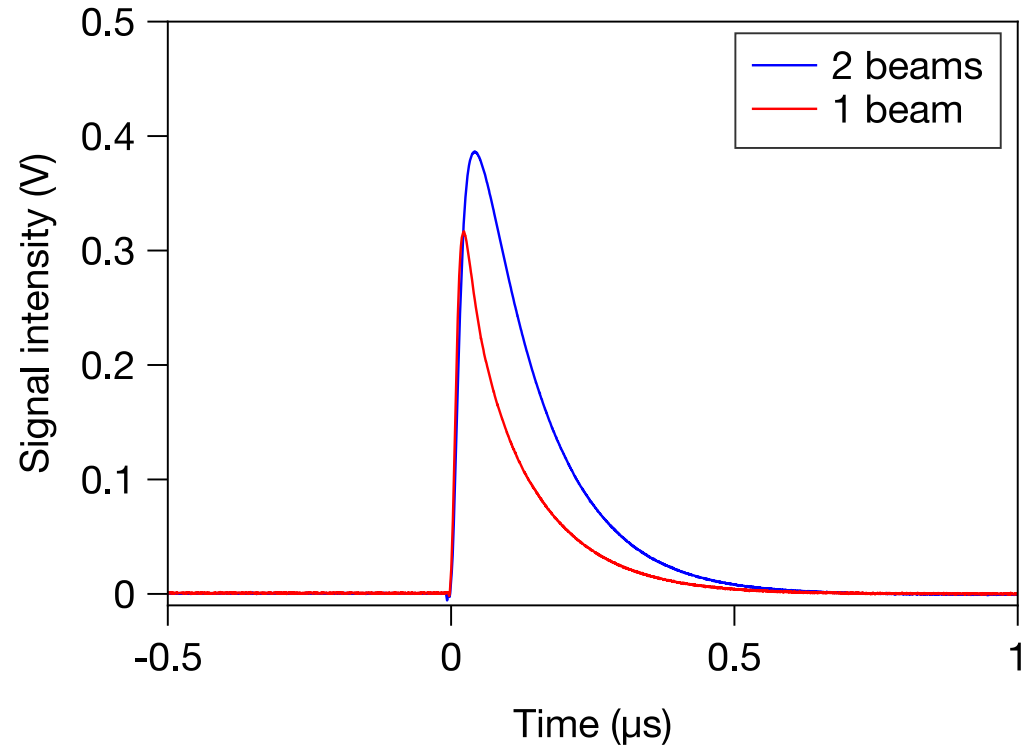

Fig. 3. Temporal waveforms of the EUV signal recorded by the energy meter for single-beam and dual-beam irradiation.

Closer inspection of the calorimeter waveforms in Fig. 3 shows that, while the peak signal amplitudes are comparable, the signal under dual-beam irradiation extends over a longer time than under single-beam irradiation. Because the calorimeter response reflects the integrated signal rather than the instantaneous EUV emission, this prolongation does not indicate a longer EUV emission time but rather an increase in the total EUV energy. The waveform duration increased from ~150 ns (FWHM) for single-pulse to ~200 ns (FWHM) for dual-pulse irradiation, corresponding to a larger integrated EUV energy. Figures 4(a) and 4(b) show the EUV pinhole images recorded under single-beam and dual-beam irradiation conditions, respectively. The EUV source sizes were 60 – 70 µm in both cases, confirming comparable source dimensions between the two irradiation schemes. Next-generation EUV lithography tools targeting high-NA and hyper-NA configurations are expected to require source sizes of approximately 100 µm or smaller. The source sizes obtained in the present work satisfy this requirement, owing to the low laser pulse energy and tight focusing employed. In both cases, the EUV source size exceeded the laser focus diameter due to plasma expansion yet remained on the same order of magnitude.

Given that the EUV source sizes are nearly identical for both configurations (60 – 70 µm; see Fig. 4), the EUV CE enhancement cannot be explained by an increase in the emitting area alone. The longer emission duration suggests that the dual-beam geometry more effectively suppresses cooling due to three-dimensional plasma expansion, thereby maintaining the plasma at the optimal electron temperature (~30 eV) for EUV emission over a longer time window. Since the total radiated EUV energy is the time integral of the emitted power, this temporal effect accounts for a substantial fraction of the observed 40% EUV CE improvement.

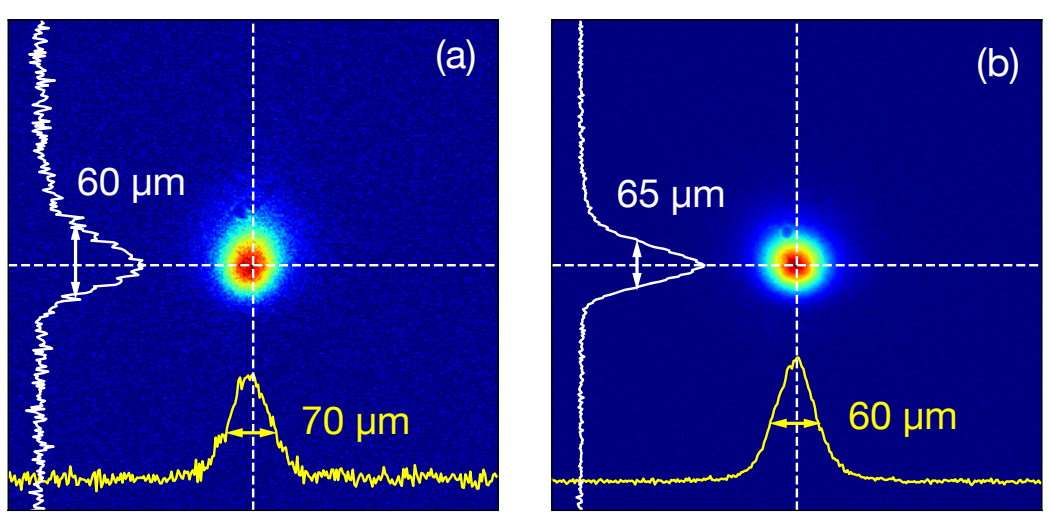

Fig. 4. EUV source images by a pinhole camera at different numbers of laser beams under a total irradiated laser energy of 40 mJ. (a) 1 beam (40 mJ/pulse, spot diameter of 40 µm), (b) 2 beams (20 mJ/pulse, spot diameter of 30 µm) at the incident angles of ± 30°, keeping each laser intensity at $1.6 \times 10^{11}$ W/cm$^2$ by changing the focal positions and focal spot diameters of the focusing lens. The laser beams irradiated to the planar Sn target simultaneously.

This enhancement can be attributed to differences in the plasma heating geometry between the two configurations. In the single-beam case, energy deposition is confined along the laser propagation axis, leaving plasma in the transverse direction weakly heated and contributing little to EUV emission. Furthermore, for the sub-40-µm focal spots used here, three-dimensional plasma expansion causes rapid cooling that shortens the EUV emission duration and limits the total EUV yield [11]. Dual-beam irradiation mitigates both the restricted heating volume and the rapid expansion cooling by effectively increasing the heated plasma volume and increasing the EUV-emitting area, thereby boosting both the emission energy and the EUV CE. These results confirm that multiple-beam irradiation schemes are beneficial for 2-µm-driven LPP EUV sources, just as with the 1-µm laser irradiation. To assess the practical applicability of this approach, it is essential to verify that the EUV source size remains within acceptable bounds under dual-beam irradiation, since an enlarged source would tighten the etendue limitation on the downstream collection optics.

The EUV emission originates primarily from the plasma surface, where the effective emissivity is maximized at an electron density on the order of $10^{19}$ – $10^{20}$ $cm^{-3}$, reflecting a balance between emission and self-absorption. Because Sn laser-produced plasmas exhibit a large optical depth for in-band EUV radiation, emission generated within the plasma core is reabsorbed before reaching the surface; only radiation emitted in a thin surface layer can escape. Consequently, the total EUV yield scales with the plasma surface area and, hence, the EUV source size. Under dual-beam irradiation, the oblique pre-pulse heated a larger volume of the low-density plasma, thereby expanding the emitting surface area and increasing the EUV output.

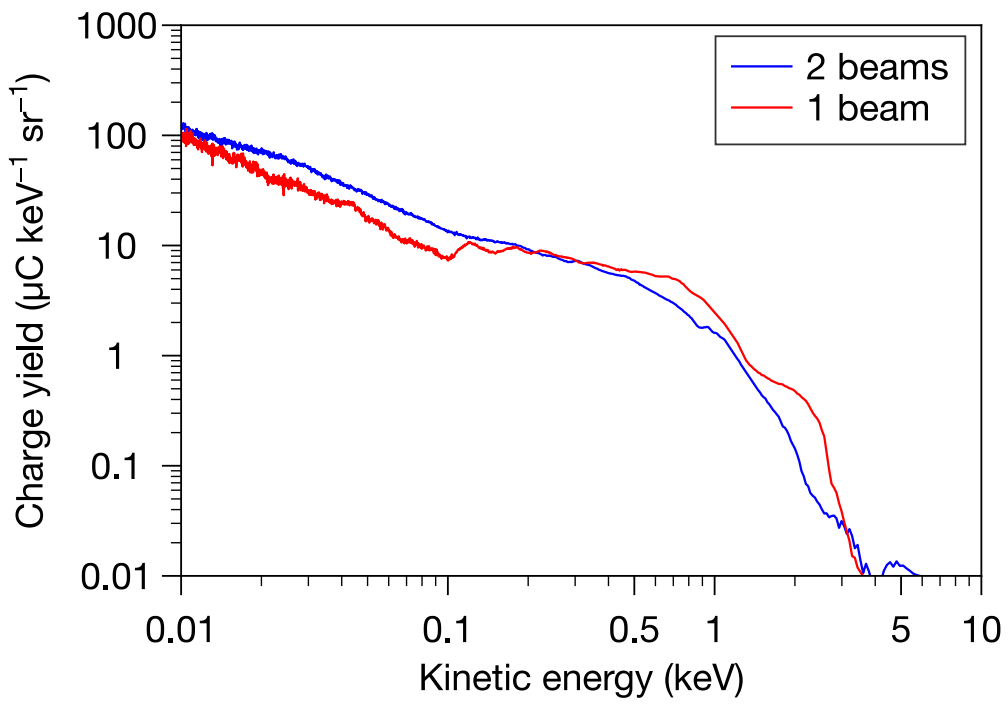

Fig. 5. Energy spectra of Sn ions by the Faraday cup for the number of beams of one or dual laser pulse irradiation conditions.

To verify that equivalent plasma conditions were maintained at a fixed total laser energy of 40 mJ, we measured energetic ion spectra for different numbers of laser pulses, as shown in Fig. 5. The energy of the emitted ions is governed by the electron temperature, initial electron density, and laser pulse duration [20,21], thus, comparable ion spectra serve as a reliable indicator of similar plasma conditions. Under dual-beam irradiation, the spectra exhibit consistent profiles and peak energies regardless of the number of pulses, confirming that equivalent plasma conditions were reproducibly generated in each case.

In summary, we demonstrated a 40% enhancement in EUV CE by simultaneous dual-beam 2-μm laser irradiation of a planar Sn target. Single-beam irradiation at 40 mJ ($1.6 \times 10^{11}$ W/$cm^2$) yielded a CE of 2.6%; splitting the same energy into two equal beams raised the CE to 3.6%, the highest reported for this target geometry. The EUV source size and energetic-ion spectra were nearly identical for both configurations, confirming comparable plasma conditions. The scheme scales straightforwardly by cascading beam splitters and requires only passive optics — making it compatible with existing 100-kHz laser architectures without added complexity. Extension to mass-limited droplet targets [3] and to three or more beams [7] is expected to yield further CE gains. This multi-beam approach is also applicable for beyond 13.5-nm lithography, and for Blue-X and water-window sources [22].

**Funding.** RIKEN TRIP initiative (Leading-edge semiconductor technology); JST K Program (No. JP MJKP24M1); Japan Society for the Promotion of Science (JSPS) (Nos. S20063, JP 23H01416, JP 24H00836, JP 25H00395); Sumitomo Foundation (No. 2200578).

**Acknowledgments.** The authors are indebted to Shuma Ako, Kimihiro Omata, and Shintaro Hayasaka (Utsunomiya University) for their helpful technical support and discussions.

**Disclosures.** The authors declare no conflicts of interest.

**Data availability.** Data underlying the results presented in this Letter are not publicly available but may be obtained from the authors upon reasonable request.